\documentclass[10pt]{article}
\usepackage{graphicx}
\usepackage{mciteplus}
\usepackage{multirow}
\usepackage{bm}
\usepackage{hyperref}
\usepackage{amsmath}

\def\Title#1{\begin{center} {\Large #1 } \end{center}}
\def\Author#1{\begin{center}{ \sc #1 } \end{center}}
\def\MultiAuthor#1#2{\sc #1$^{[#2]}$,~}
\def\SpeakerAuthor#1#2{\sc #1$^{[#2]}$\footnote{\it Speaker},~}
\def\LastAuthor#1#2{and \sc #1$^{[#2]}$}
\def\Address#1{\begin{center}{ \it #1} \end{center}}

\newcommand\pubblock{\rightline{\begin{tabular}{l} Proceedings of the Fifth Annual LHCP\\ \pubnumber\\
         \pubdate  \end{tabular}}}

\newenvironment{Abstract}{\begin{quotation} \begin{center} 
             \large ABSTRACT \end{center}\bigskip 
      \begin{center}\begin{large}}{\end{large}\end{center} \end{quotation}}

\newenvironment{Presented}{\begin{quotation} \begin{center} 
             PRESENTED AT\end{center}\bigskip 
      \begin{center}\begin{large}}{\end{large}\end{center} \end{quotation}}

\def\beq{\begin{equation}}
\def\eeq#1{\label{#1}\end{equation}}
\def\eeqn{\end{equation}}

\def\beqa{\begin{eqnarray}}
\def\eeqa#1{\label{#1}\end{eqnarray}}
\def\eeqan{\end{eqnarray}}

\let\bar=\overbar

\def\Dslash{\not{\hbox{\kern-4pt $D$}}}
\def\dslash{\not{\hbox{\kern-2pt $\del$}}}

\def\msb{{\bar{\ssstyle M \kern -1pt S}}}

\usepackage{color}

\newcommand\pubnumber{arXiv:1710.09644 [hep-ph]}

\newcommand\pubdate{\today}

\newcommand{\dmd}{\Delta m_d}
\newcommand{\dms}{\Delta m_s}

\newcommand{\rhob}{\bar \rho}
\newcommand{\etab}{\bar \eta}
\newcommand{\vubsvcb}{\left | V_{ub}/V_{cb}  \right |}
\newcommand{\vubbf}{${\textbf{\bf{V}}_{\textbf{\footnotesize\bf{ub}}}}$}
\newcommand{\vcbbf}{${\textbf{\bf{V}}_{\textbf{\footnotesize\bf{cb}}}}$}

\def\utfit{{\bf{U}}\kern-.20em{\bf{T}}\kern-.15em{\it{fit}}\@}
\def\butfit{{\bf{U}}\kern-.18em{\bf{T}}\kern-.10em{\it{fit}}\@}

\begin{document}

\large
\begin{titlepage}
\pubblock

\vfill
\Title{ Unitarity Triangle Analysis in the Standard Model and Beyond }
\vfill

\Author{\utfit\ Collaboration:}
\begin{center}
\MultiAuthor{Cristiano Alpigiani}{a} 
\MultiAuthor{Adrian Bevan}{b}
\SpeakerAuthor{Marcella Bona}{b}
\MultiAuthor{Marco Ciuchini}{c}
\MultiAuthor{Denis Derkach}{d}
\MultiAuthor{Enrico Franco}{e}
\MultiAuthor{Vittorio Lubicz}{f}
\MultiAuthor{Guido Martinelli}{g}
\MultiAuthor{Fabrizio Parodi}{h}
\MultiAuthor{Maurizio Pierini}{i}
\MultiAuthor{Luca Silvestrini}{e}
\MultiAuthor{Viola Sordini}{j}
\MultiAuthor{Achille Stocchi}{k}
\MultiAuthor{Cecilia Tarantino}{f}
\LastAuthor{Vincenzo Vagnoni}{l}
\end{center}
\Address{$^{[a]}$University of Washington,
$^{[b]}$Queen Mary University of London,
$^{[c]}$INFN Sezione di Roma Tre,
$^{[d]}$Yandex/Higher School of Economics,
$^{[e]}$INFN Sezione di Roma,
$^{[f]}$University of Roma Tre,
$^{[g]}$University of Roma La Sapienza,
$^{[h]}$University of Genova and INFN,
$^{[i]}$CERN,
$^{[j]}$IPNL-IN2P3 Lyon,
$^{[k]}$LAL-IN2P3 Orsay,
$^{[l]}$INFN Sezione di Bologna.
}
\vfill
\begin{Abstract}

  Flavour physics represents a unique test bench for the Standard Model (SM).
  New analyses performed at the LHC experiments are now providing unprecedented
  insights into CKM metrology and new evidences for rare decays. The CKM picture
  can provide very precise SM predictions through global analyses.
  We present here the results of the latest global SM analysis performed by the
  \utfit\ collaboration including all the most updated inputs from experiments,
  lattice QCD and phenomenological calculations.
  In addition, the Unitarity Triangle (UT) analysis can be used to constrain the
  parameter space in possible new physics (NP) scenarios. We update here also
  the UT analysis beyond the SM by the \utfit\ collaboration.
  All of the available experimental and theoretical information on
  $\Delta F=2$ processes is reinterpreted including a model-independent NP parametrisation.
  We determine the allowed NP contributions in the kaon, $D$, $B_d$, and $B_s$
  sectors and, in various NP scenarios, we translate them into bounds for the NP scale
  as a function of NP couplings.

\end{Abstract}
\vfill

\begin{Presented}
The Fifth Annual Conference\\
on Large Hadron Collider Physics \\
Shanghai Jiao Tong University, Shanghai, China\\ 
May 15-20, 2017
\end{Presented}
\vfill
\end{titlepage}
\def\thefootnote{\fnsymbol{footnote}}
\setcounter{footnote}{0}
%

\normalsize 


\section{Introduction}

Flavour physics represents a powerful tool to test the Standard Model (SM),
to quantify the coherence of its picture and to explore possible departures
from it. From the flavour global fit we can extract the most accurate
determination of the parameters of the CKM matrix~\cite{cabibbo,*KM},
as well as the best SM predictions of flavour observables.
The Unitarity Triangle (UT) analysis here presented is performed by the
\utfit~Collaboration following the method described in
Refs.~\cite{Ciuchini:2000de,Bona:2005vz}.
We updated the analysis with the latest determinations of the theoretical
inputs and the latest measurements of the experimental observables.
The basic constraints used in the global fit and contributing to the
sensitivity of the CKM matrix elements are: $\vubsvcb$ from semileptonic
$B$ decays, $\dmd$ and $\dms$ from $B^0_{d,s}$ oscillations,
$\varepsilon_K$ from neutral $K$ mixing, $\alpha$ UT angle from charmless
hadronic $B$ decays, $\gamma$ UT angle from charm hadronic $B$ decays,
and the sine of $2\beta$ UT angle from $B^0\to J/\psi K^0$ decays.

The values of most experimental inputs are taken from the Heavy Flavour Averaging
Group (HFLAV)~\cite{hflav}, however when most updated individual results are available
the \utfit\ collaboration performs its own averages.
Below a specific update is discussed for the $\vubsvcb$ experimental input.
On the theoretical side, the non-perturbative QCD parameters are taken
from the most recent lattice QCD determinations: as a general
prescription, we average the $N_f=2+1+1$ and $N_f=2+1$ FLAG
numbers~\cite{flag16}, using eq.~(28) in Ref.~\cite{Carrasco:2014cwa}
and including the results in Ref.~\cite{Bazavov:2016nty}.
The continuously updated set of numerical values used as inputs can be found
at URL {\tt{http://www.utfit.org/}}.

\begin{figure}[!b]
  \vspace*{-0.6cm}
  \hspace*{-0.4cm}
  \centering
  \begin{tabular}{cc}
    \includegraphics[width=0.36\linewidth]{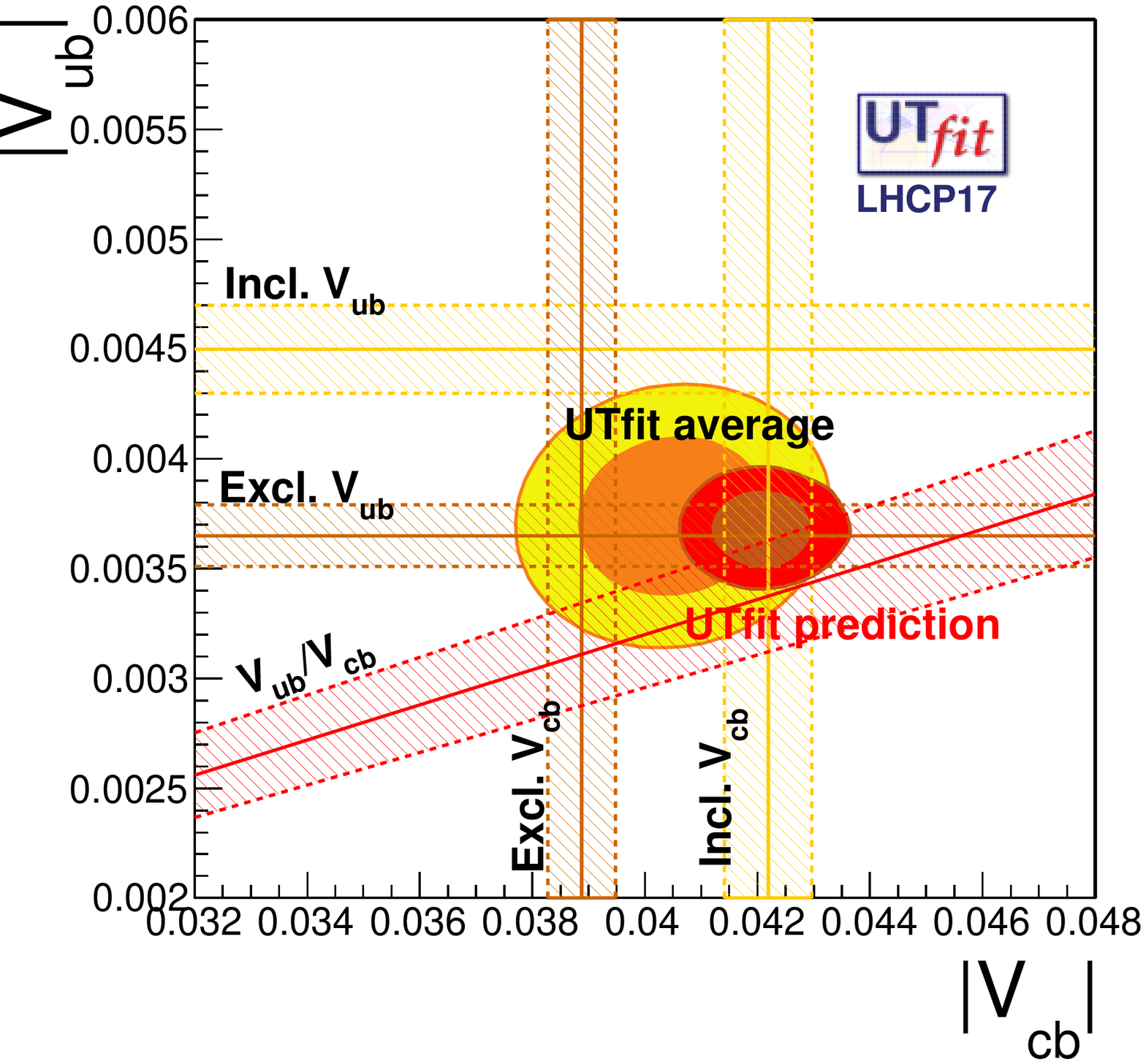} &
    \hspace*{-1.0cm}
    \includegraphics[width=0.36\linewidth]{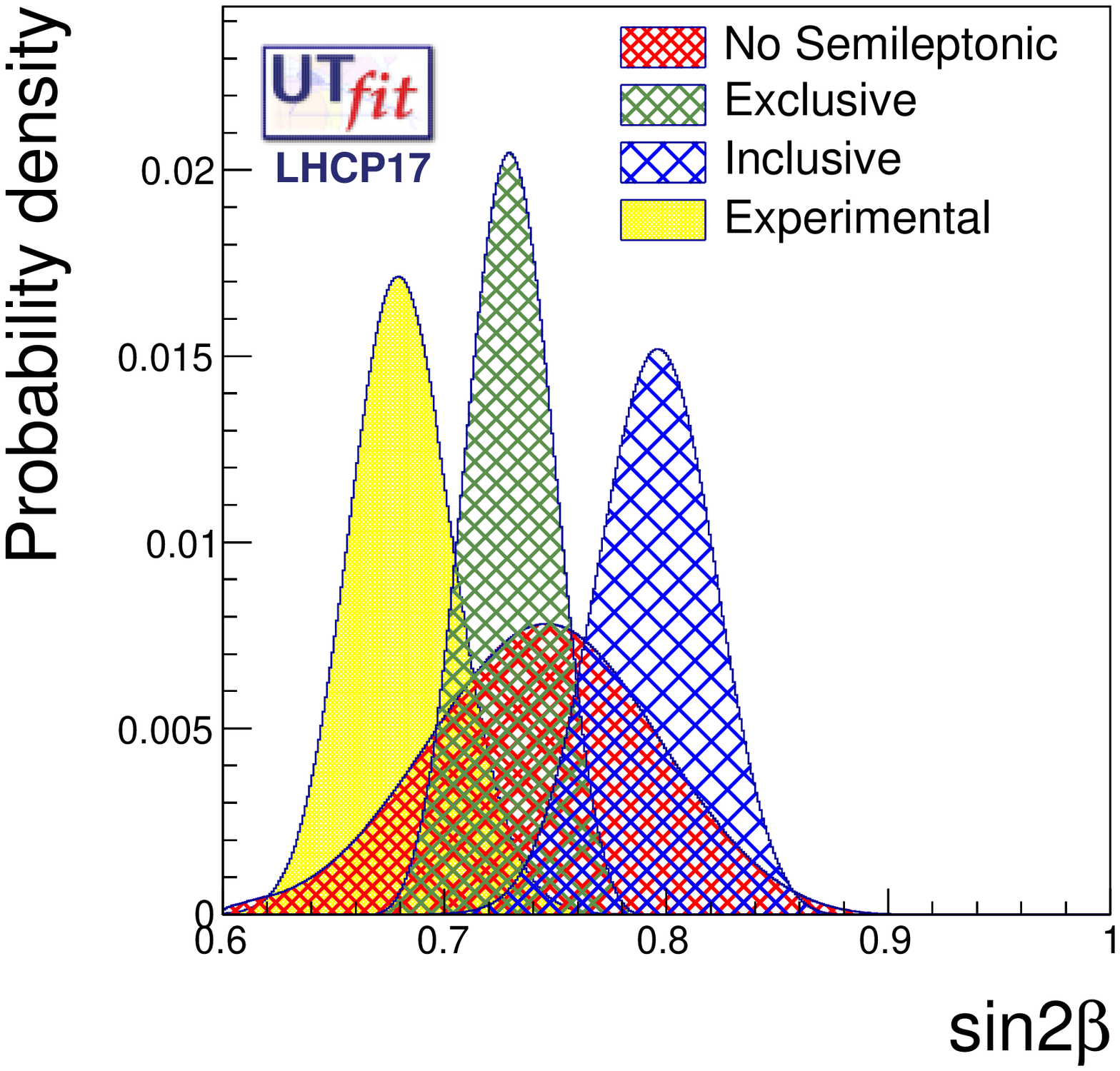} 
    \hspace*{-0.6cm}
  \end{tabular}
  \vspace*{-0.2cm}
  \caption{{\it{Left}}: $|V_{cb}|$ vs $|V_{ub}|$ plane showing
    the values reported in table~\ref{tab:vubvcb}.
    We include in the average the LHCb ratio measurement~\cite{Aaij:2015bfa}
    that is shown as a diagonal band.
    {\it{Right}}: predictions on $\sin 2\beta$ from the SM global fits
    obtained when changing the inputs as indicated in the legend.
  }
  \label{fig:vubvcb}
\end{figure}

\begin{figure}[!t]
  \vspace*{-0.6cm}
  \hspace*{-0.4cm}
  \centering
  \begin{tabular}{cc}
    \includegraphics[width=0.38\linewidth]{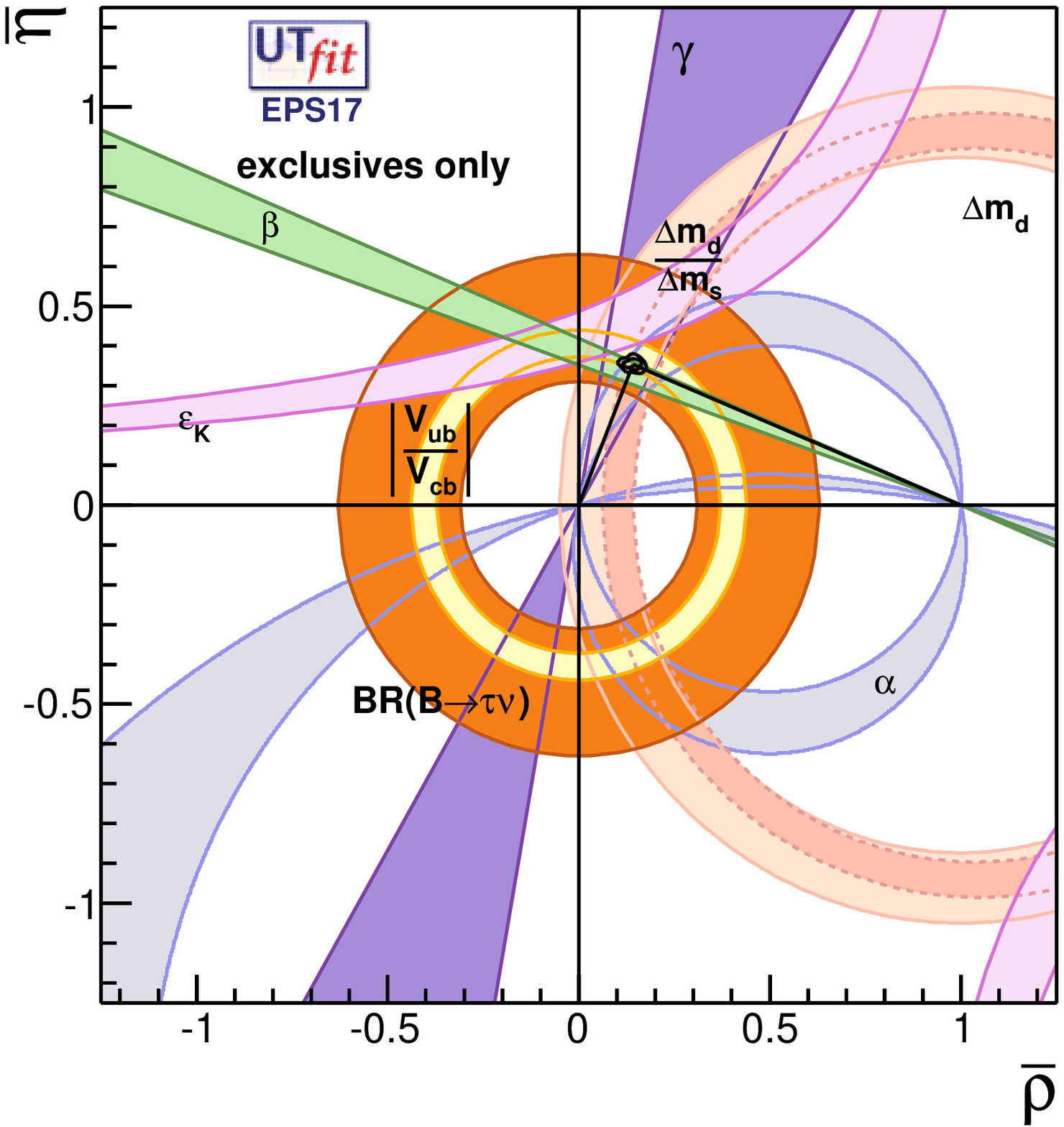} &
    \hspace*{-1.0cm}
    \includegraphics[width=0.38\linewidth]{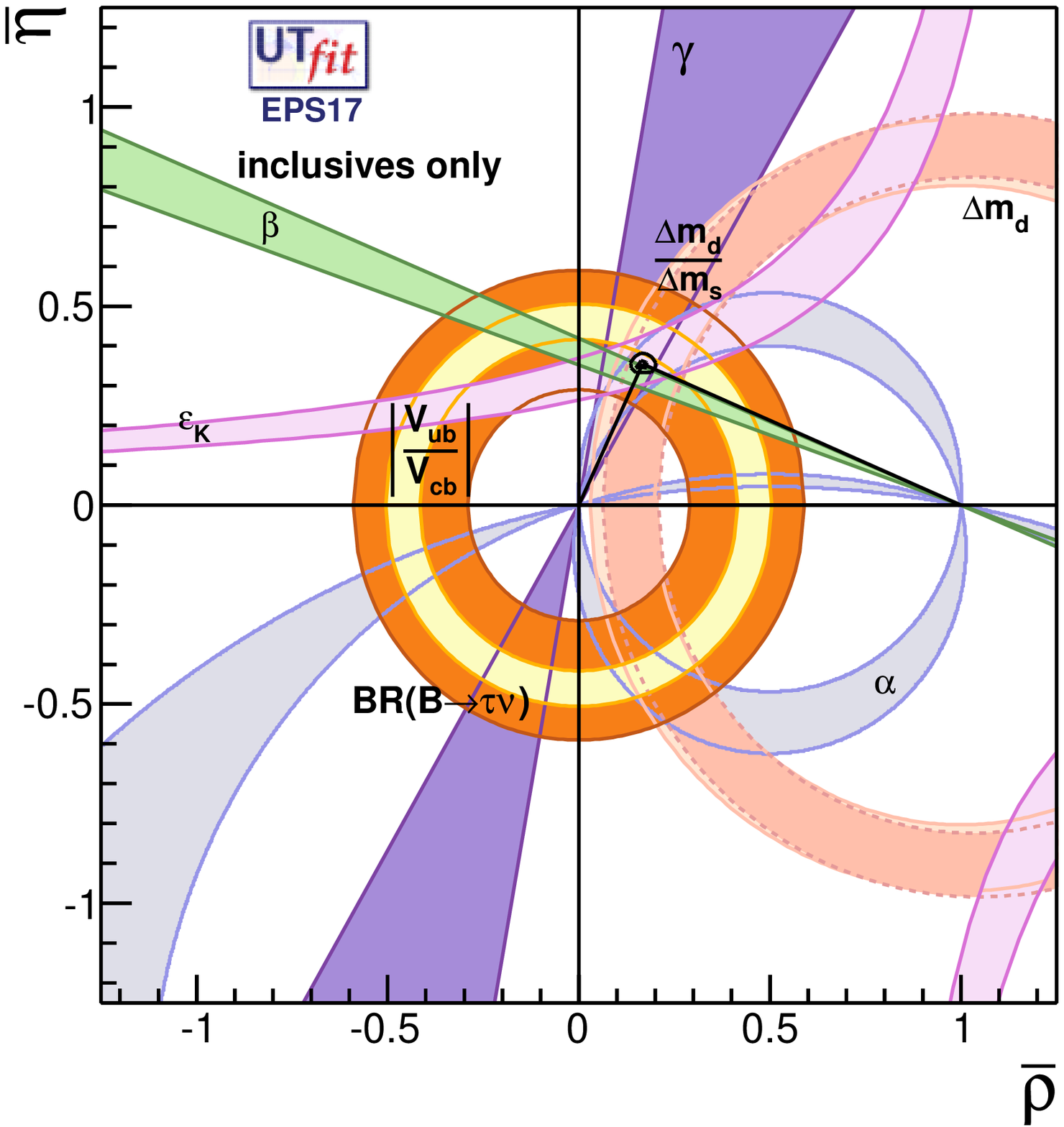}
    \hspace*{-0.6cm}
  \end{tabular}
  \vspace*{-0.2cm}
  \caption{$\bar \rho$-$\bar \eta$ plane with the SM global fit results
    using only exclusive inputs for both $V_{ub}$ and $V_{cb}$
    ({\it{left}}) and using only inclusive inputs ({\it{right}}).
    The black contours display the 68\% and 95\% probability
    regions selected by the given global fit. The 95\% probability
    regions selected by the single constraints are also shown.
  }
  \label{fig:rhoetainex}
\end{figure}

\section{Updated inputs and results of the global fit in the SM}
\label{sec:input}

\begin{table}[!h]
  \vspace*{-0.5cm}
  \centering
  \vspace*{0.1cm}
  \begin{tabular}{ccc|c|c}
    [$10^{-3}$] & {\bf{excl.}}~~~~~~& {\bf{incl.}}~~~~~~& $|$\vubbf$/|$\vcbbf$|$ & {\bf{2D average}} \\
    \hline
    $|$\vcbbf$|$   & $38.88 \pm 0.60$  & $42.19 \pm 0.78$  & \multirow{ 2}{*}{$(8.0 \pm 0.6) 10^{-2}$} & $40.5 \pm 1.1$ \\
    $|$\vubbf$|$   & $3.65 \pm 0.14$   & $4.50 \pm 0.20$   &                                           & $3.74 \pm 0.23$
  \end{tabular}
  \caption{$V_{cb}$ and $V_{ub}$ experimental inputs are shown as
    values. The individual $V_{cb}$ and $V_{ub}$ exclusive and
    inclusive numbers are taken from the most updated HFLAV averages~\cite{hflav}.
  }
  \label{tab:vubvcb}
  \vspace*{-0.1cm}
\end{table}

For the inputs coming from the semileptonic $B$ decays, we use the
values shown in the left plot in Fig.~\ref{fig:vubvcb} and listed in Table~\ref{tab:vubvcb}.
The \utfit\ two-dimensional (2D) average shown is calculated with a 2D procedure
inspired by the skeptical method of Ref.~\cite{DAgostini:1999niu} with $\sigma=1$.
A very similar result is obtained from a 2D {\it{\`a la}} PDG~\cite{pdg16}
procedure.
It is evident that exclusive and inclusive results persist in showing discrepancies
at the level of about $3.3\sigma$ in the case of $V_{cb}$ and about $3.4\sigma$ for $V_{ub}$.
The effect of these deviations in the global fit results are shown in Figs.~\ref{fig:vubvcb}
and~\ref{fig:rhoetainex}.
The right plot in Fig.~\ref{fig:vubvcb} shows the predictions for $\sin 2\beta$
from the SM global fits obtained when changing the inputs relative to the
semileptonic $B$ decays, using only exclusive inputs for both $V_{ub}$ and $V_{cb}$,
using only inclusive inputs or not using the $V_{ub}$ and $V_{cb}$ inputs
at all. The experimental value for $\sin 2\beta$ is also shown for comparison.
Fig.~\ref{fig:rhoetainex} shows the global fit results on the $\rhob-\etab$ plane 
using only exclusive inputs for both $V_{ub}$ and $V_{cb}$ or using only inclusive inputs.
These inclusive-vs-exclusive discrepancies have been highlighted
and discussed by the \utfit~collaboration since $2006$~\cite{Bona:2006ah}.

The angle $\gamma$ of the UT is extracted from $V_{cb}$ and $V_{ub}$
mediated transitions in $B \rightarrow D^{(*)}K^{(*)}$ decays.
The current \utfit\ input value for $\gamma$ is $(69.8 \pm 5.9)^\circ$,
updated after the winter 2017 experimental results.
The angle $\alpha$ of the UT is obtained from charmless two-body $B$ decays
in $\pi\pi$, $\rho\rho$ or $\rho\pi$ final states via isospin analyses.
The current \utfit\ value of the SM solution corresponds
to $\alpha_{SM} = (94.2 \pm 4.5)^\circ$~\cite{UTalpha,bona-ichep16-sm}.

\begin{figure}[!bt]
  \vspace*{-0.2cm}
  \hspace*{-0.4cm}
  \centering
  \begin{tabular}{cc}
    \includegraphics[width=0.38\linewidth]{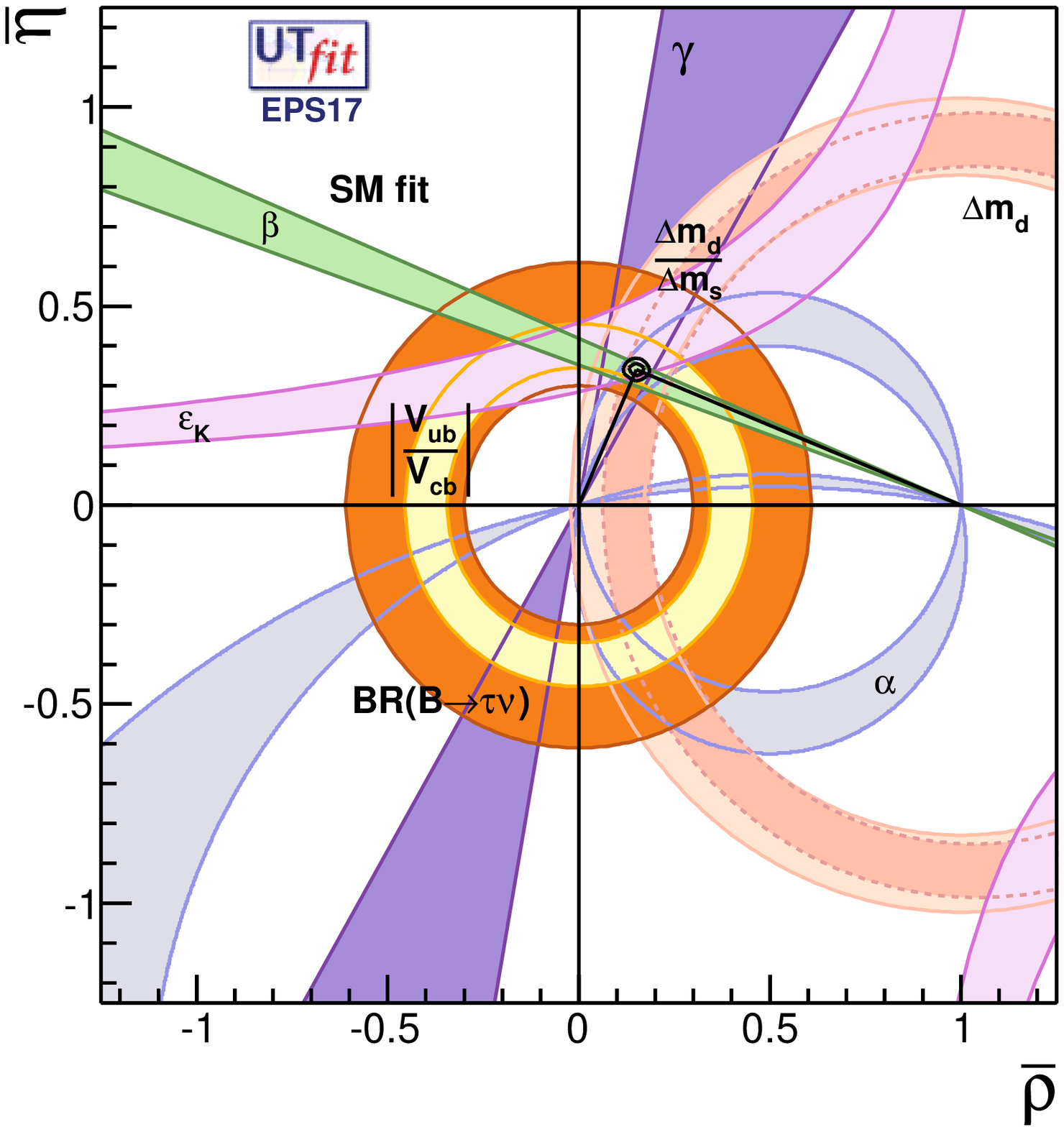} &
    \hspace*{-1.0cm}
    \includegraphics[width=0.38\linewidth]{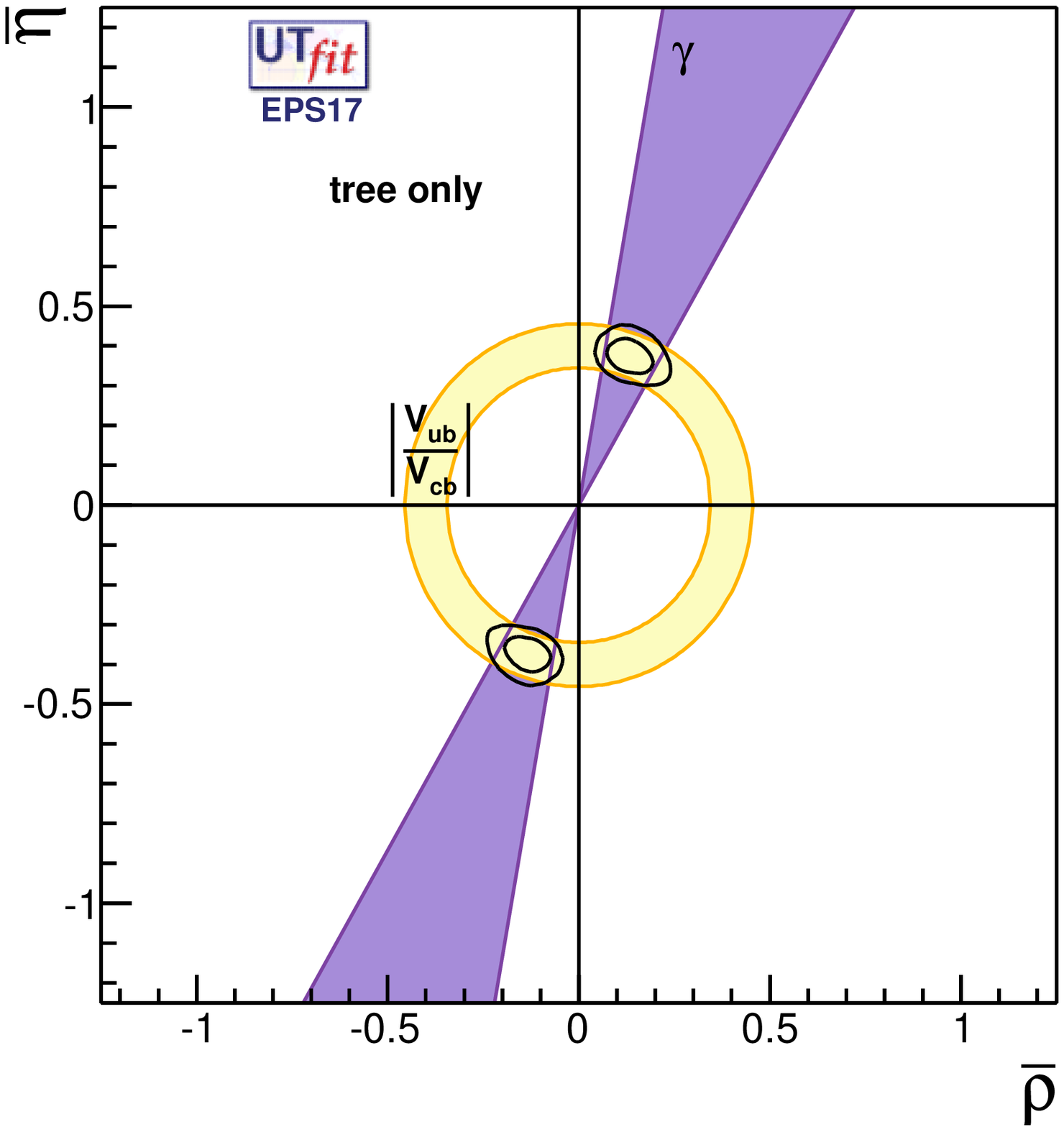}
    \hspace*{-0.6cm}
  \end{tabular}
  \vspace*{-0.2cm}
  \caption{{\it{Left:}} $\rhob-\etab$ plane showing the result of the SM fit.
    {\it{Right:}} $\rhob-\etab$ plane showing the result of the tree-only fit.
    The black contours display the 68\% and 95\% probability
    regions selected by the given global fit. The 95\% probability
    regions selected by the single constraints are also shown.}
  \label{fig:smtree}
\end{figure}

\begin{figure}[!bt]
  \vspace*{-0.6cm}
  \hspace*{-0.4cm}
  \centering
  \begin{tabular}{cc}
    \includegraphics[width=0.38\linewidth]{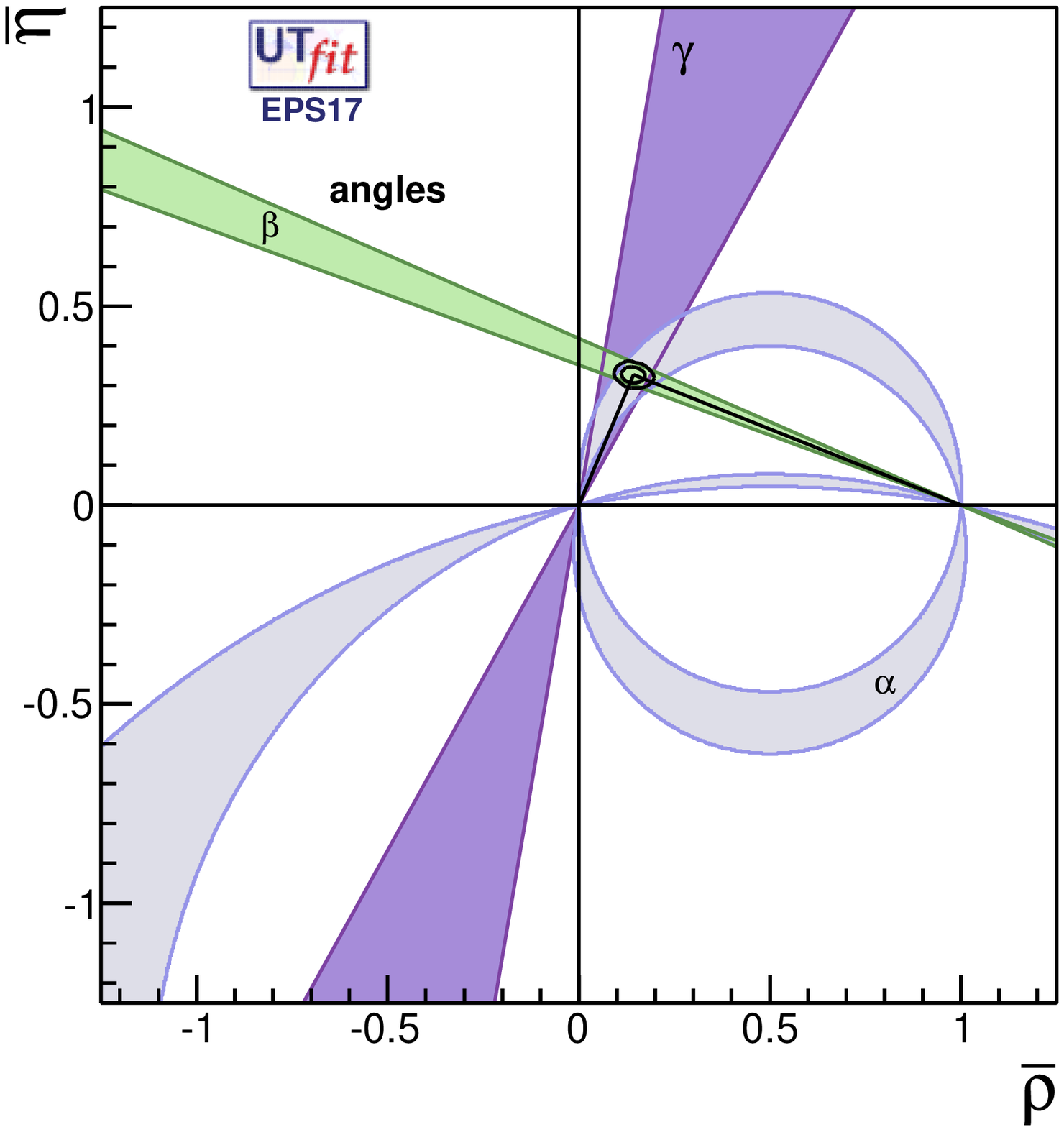} &
    \hspace*{-1.0cm}
    \includegraphics[width=0.38\linewidth]{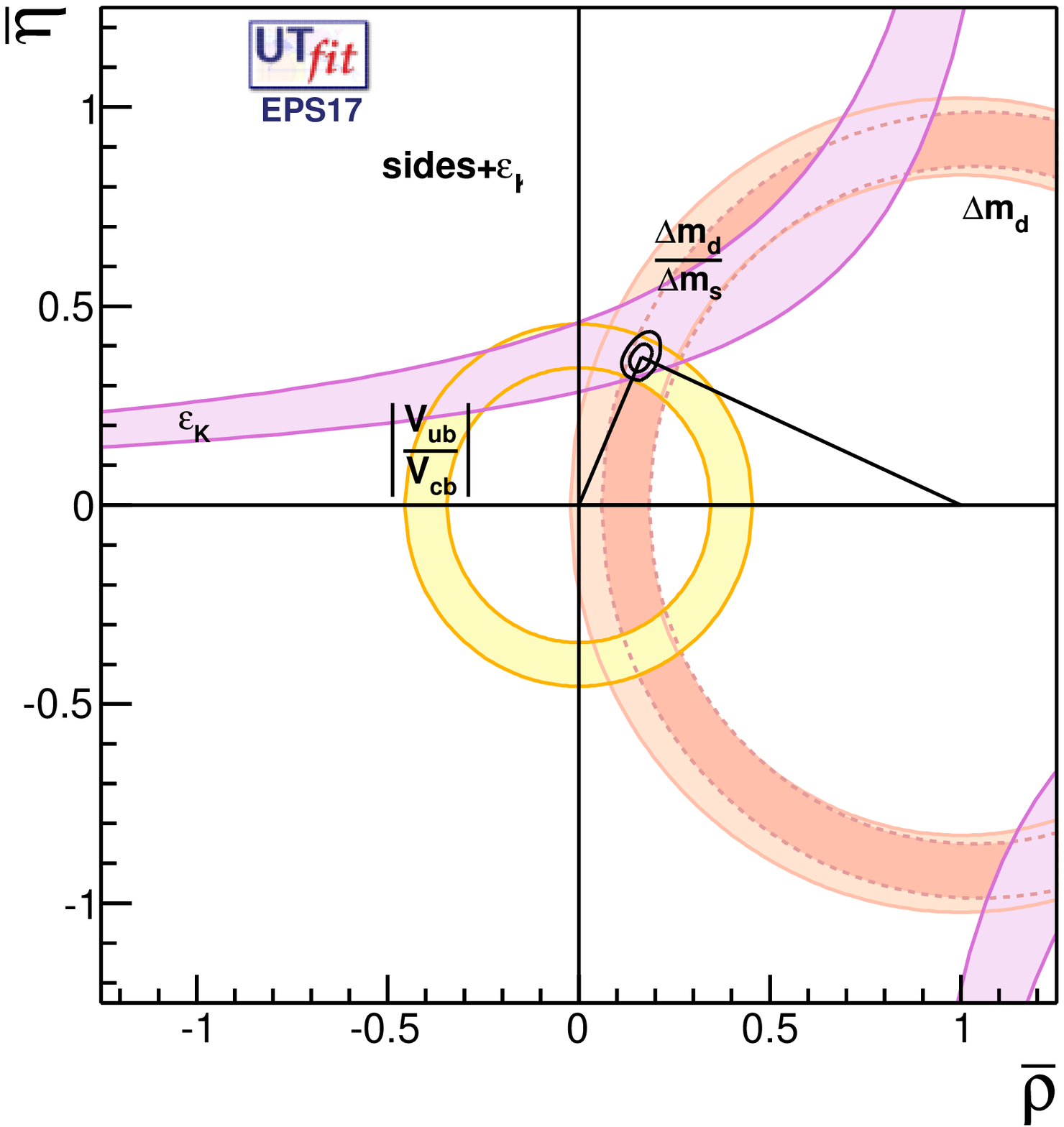}
    \hspace*{-0.6cm}
  \end{tabular}
  \vspace*{-0.2cm}
  \caption{{\it{Left:}} $\rhob-\etab$ plane showing the result of the SM fit
    using only the UT angles as constraints.
    {\it{Right:}} $\rhob-\etab$ plane showing the result of the SM fit
    using the UT sides and the kaon mixing as constraints.
    The black contours display the 68\% and 95\% probability
    regions selected by the given global fit. The 95\% probability
    regions selected by the single constraints are also shown.}
  \label{fig:anglesides}
\end{figure}

Using the latest inputs and our Bayesian framework, we perform
the global fit to extract the CKM matrix parameters $\bar\rho$
and $\bar\eta$: we obtain $\bar \rho=0.151 \pm 0.014$ and
$\bar \eta=0.342 \pm 0.013$. The left plot in Fig.~\ref{fig:smtree} shows the
result of the SM fit on the $\bar\rho$-$\bar\eta$ plane, while the right
figure shows the ``tree-only'' fit when only tree-level measurements
are included ($\vubsvcb$ and $\gamma$ assumed NP-free).

We also perform our fit separating two sets of inputs as shown in
Figure~\ref{fig:anglesides}: the ``angle-only'' fit using as constraints
$\beta$, $\alpha$ and $\gamma$ measurements and the ``sides-and-$\varepsilon_K$''
fit using $\vubsvcb$, $\dmd$, $\dms$, and $\varepsilon_K$.
The comparison between the $\bar\rho$ and $\bar\eta$ values extracted
from these two fit configurations represents the current tension of the
SM inputs. From the ``angle-only'' fit we obtain $\bar \rho=0.143\pm0.022$ and
$\bar \eta=0.327\pm0.015$. From the ``sides-and-$\varepsilon_K$'
fit we obtain $\bar \rho=0.164\pm0.020$ and $\bar \eta=0.372\pm0.025$.

The main tension still present in the global fit comes from the
inclusive-vs-exclusive values of the semileptonic determinations:
for example, the inclusive $|V_{ub}|$ value shows a $\sim 3.8 \sigma$
discrepancy with respect to the rest of the fit.

\section{Result of the global fit beyond the SM}
\label{sec:np}

We now consider the UT analysis performed reinterpreting the
experimental observables including possible model-independent NP contributions.
The NP effects considered here are those entering the neutral meson mixing
($\Delta F=2$ transitions). They can be parameterised in a general way as:
\vspace*{-0.1cm}
\begin{eqnarray}
  \frac{\langle
    B_q|H_\mathrm{eff}^\mathrm{full}|\bar{B}_q\rangle} {\langle
    B_q|H_\mathrm{eff}^\mathrm{SM}|\bar{B}_q\rangle}
  \; = \; \left(1+\frac{A_q^\mathrm{NP}}{A_q^\mathrm{SM}}
  e^{2 i (\phi_q^\mathrm{NP}-\phi_q^\mathrm{SM})}\right) \nonumber
  \label{eq:nppar}
\end{eqnarray}
where in the SM it is $A_q^\mathrm{NP}=0$ and $\phi_q^\mathrm{NP}=0$.
$H_\mathrm{eff}^\mathrm{SM}$ is the SM $\Delta F=2$ effective Hamiltonian,
while $H_\mathrm{eff}^\mathrm{full}$ is its extension in a general NP model,
and $q=d$ or $s$.
The following experimental inputs are added to the NP global fit to extract
information on the $B_s$ system: the semileptonic asymmetry in $B_d$ and
$B_s$ decays, the di-muon charge asymmetry, the $B_s$ lifetime from
flavour-specific final states, and the CP-violating phase and the decay-width
difference for $B_s$ mesons from the time-dependent angular analysis of
$B_s\to J/\psi \phi$ decays. The values used as inputs are taken
from the HFLAV~\cite{hflav} averages.

\begin{figure}[!h]
  \vspace*{-0.4cm}
  \hspace*{-1.0cm}
  \centering
  \begin{tabular}{ccc}
    \vspace*{-0.2cm}
    \includegraphics[width=0.38\linewidth]{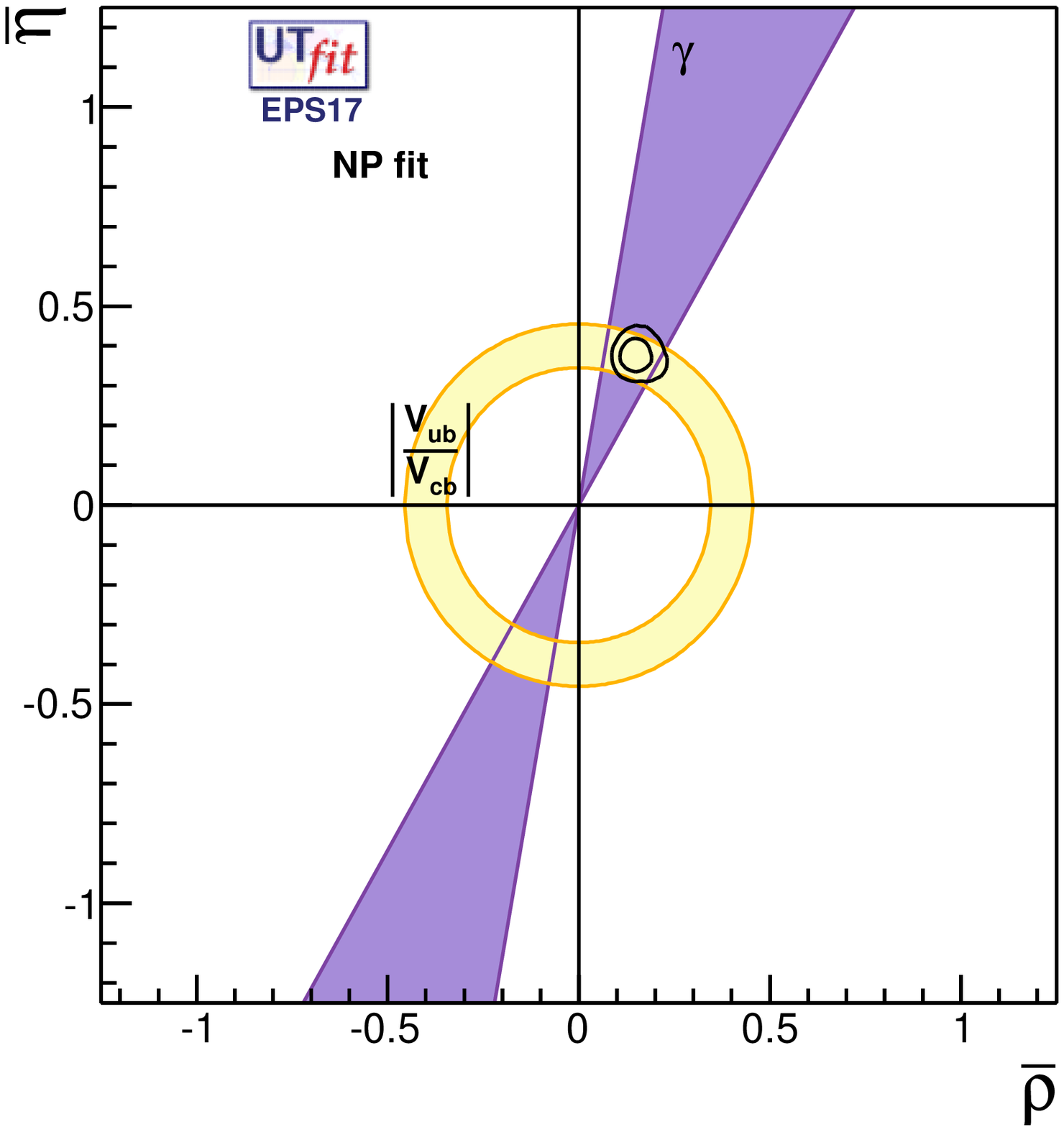} &
    \hspace*{-0.8cm}
    \includegraphics[width=0.36\linewidth]{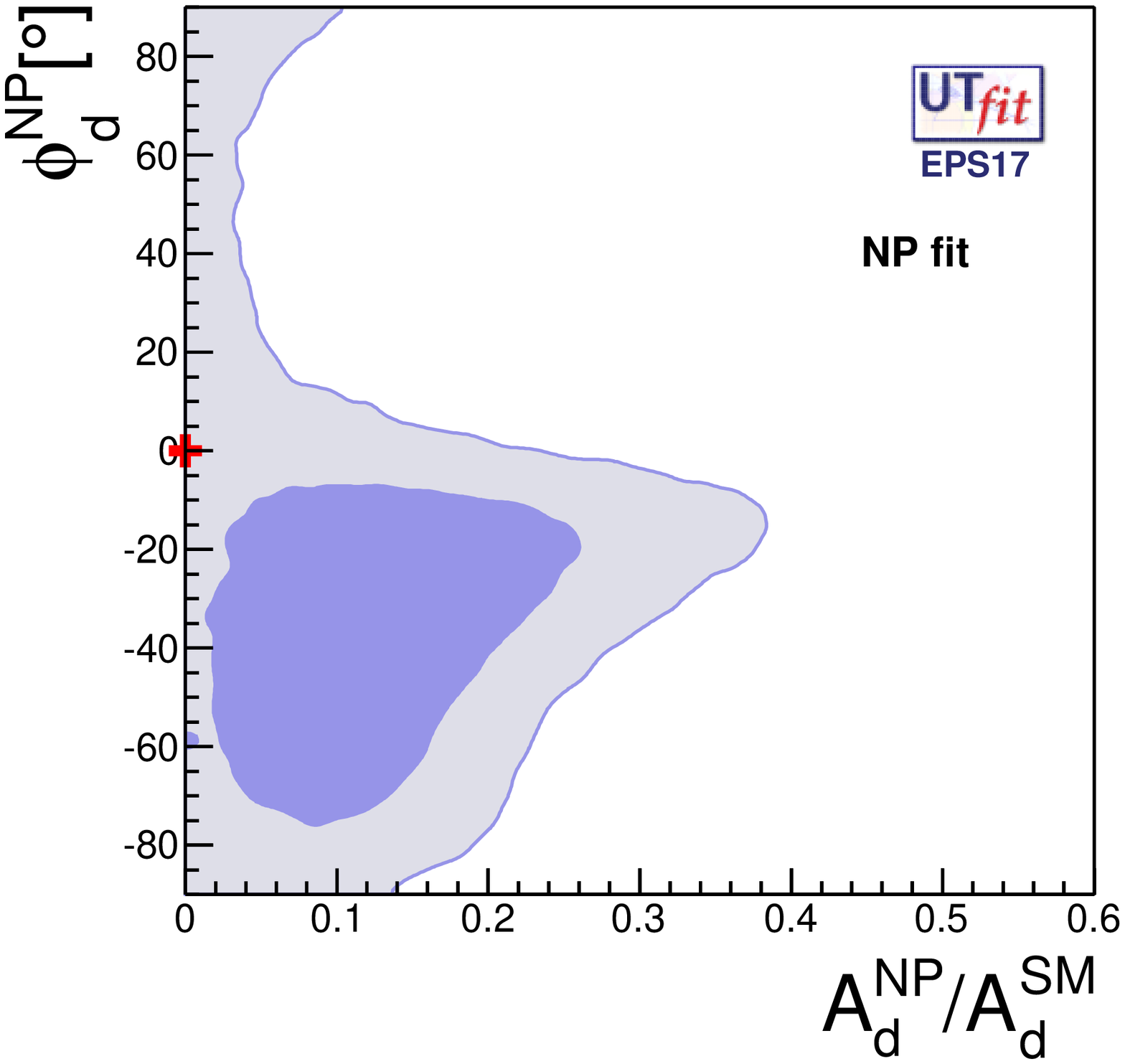} &
    \hspace*{-0.8cm}
    \includegraphics[width=0.36\linewidth]{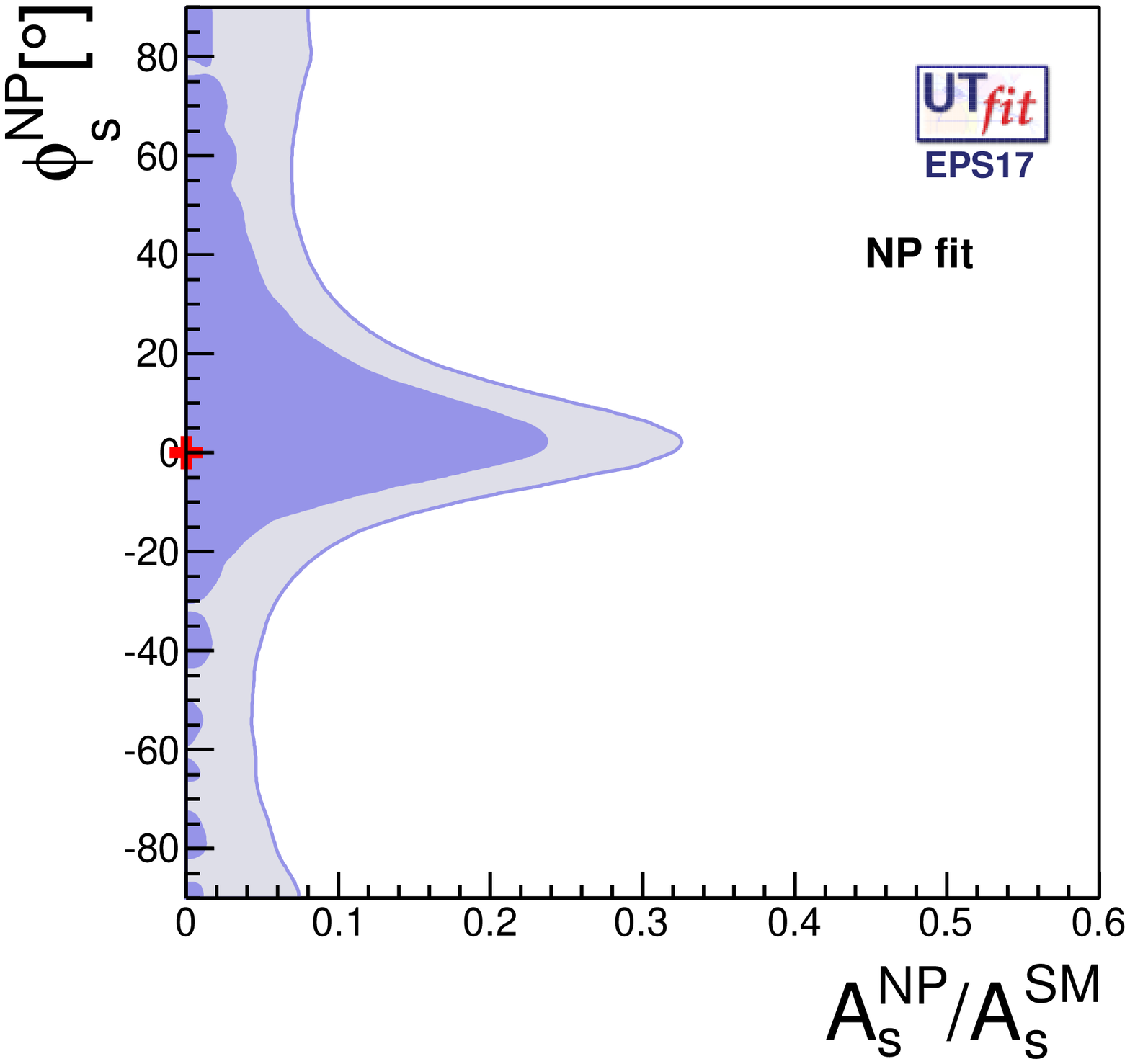}
    \hspace*{-0.6cm}
  \end{tabular}
  \caption{NP parameters in the $B_d$ system ({\it{left}}) and
    $B_s$ system ({\it{right}}), where $68\%$ (dark) and $95\%$ (light)
    probability regions are shown in the $A_q^\mathrm{NP}/A_q^\mathrm{SM}$
    -- $\phi_q^\mathrm{NP}$ planes.
    The red cross represents the SM expectation.}
  \label{fig:achille}
\end{figure}

Using the latest inputs, we perform the full NP analysis and the result of this
global fit selects a region of the $(\rhob, \etab)$ plane which is consistent with
the result of the SM analysis. This can be seen in the left $\rhob-\etab$
plot in Fig.~\ref{fig:achille}. The $\rhob$ and $\etab$ values extracted
from the NP global fit are $\rhob=0.154 \pm 0.029$ and $\etab=0.377 \pm 0.029$.
Simultaneously, the NP parameters are extracted and their allowed
ranges are shown in the two right plots in Fig.~\ref{fig:achille}.
The current tension of the SM picture is reflected in the $B_d$ sector.
In general a $30-40\%$ effect is allowed at $95\%$ probability,
given the current sensitivities.

If we now consider the most general effective Hamiltonian for
$\Delta F=2$ processes ($H_\mathrm{eff}^{\Delta F=2}$),
we can translate the current constraints into allowed ranges
for the Wilson coefficients of $H_\mathrm{eff}^{\Delta F=2}$.
The full procedure and analysis details are given in~\cite{Bona:2007vi}.
These coefficients have the general form
\vspace*{-0.1cm}
\begin{equation*}
  C_i (\Lambda) = \frac{F_i\;L_i}{\Lambda^2}\,
\end{equation*}
where $F_i$ is a function of the (complex) NP flavour couplings, $L_i$
is a loop factor that is present in models with no tree-level Flavour
Changing Neutral Currents, and $\Lambda$ is the scale of NP,
\emph{i.e.} the typical mass of the new particles mediating
$\Delta F=2$ transitions.
For a generic strongly-interacting theory with arbitrary flavour structure,
one expects $F_i \sim L_i \sim 1$ so that the allowed range from the fit
for each of the $C_i(\Lambda)$ can be immediately translated into a lower
bound on $\Lambda$. Specific assumptions on the flavour structure of NP,
for example Next-to-Minimal~\cite{papucci} Flavour Violation (NMFV),
correspond to particular choices of the $F_i$ functions.
In the case of NMFV, we have $\vert F_i\vert = F_\mathrm{SM}$
with an arbitrary phase~\cite{papucci}.
To obtain the lower bound on $\Lambda$ for loop-mediated
contributions, one simply multiplies the bounds we quote in the
following by $\alpha_s\sim 0.1$ or by $\alpha_W \sim 0.03$.

\begin{figure}[!tb]
  \centering
  \vspace*{-0.4cm}
  \hspace*{-0.2cm}
  \includegraphics[width=0.46\linewidth]{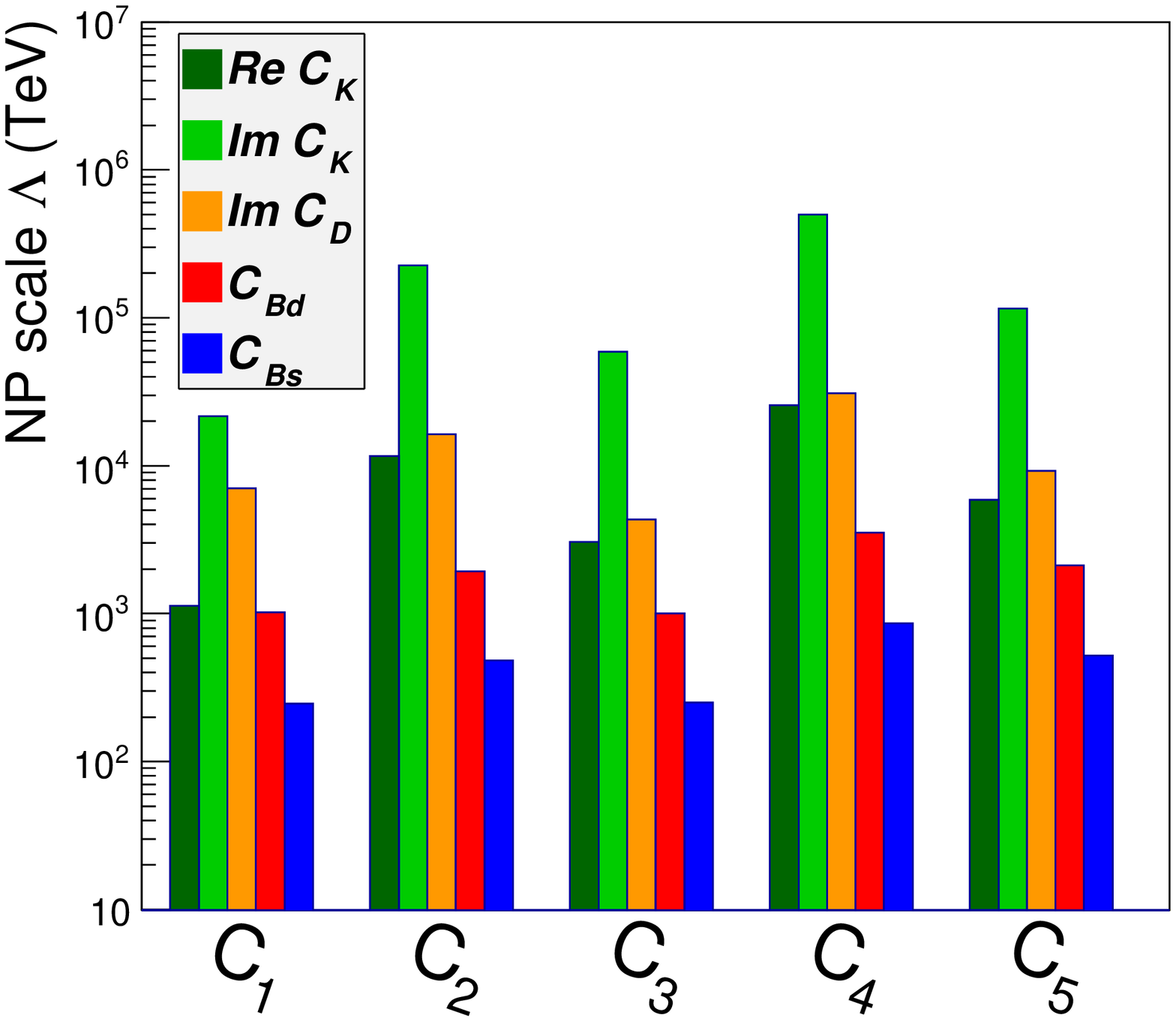}
  \includegraphics[width=0.46\linewidth]{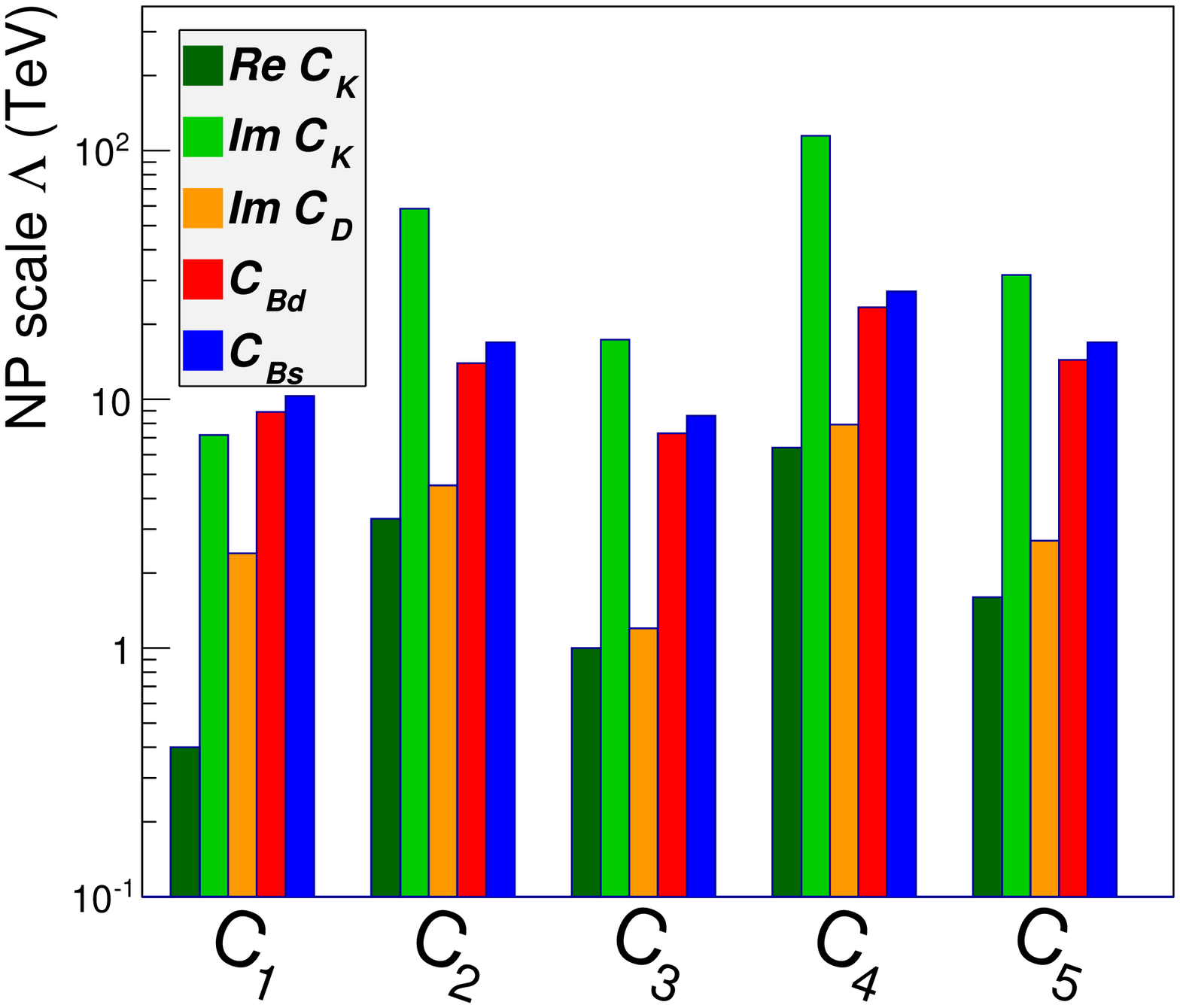}
  \hspace*{-0.2cm}
  \caption{Summary of the $95\%$ probability lower bound on the NP
    scale $\Lambda$ for strongly-interacting NP in the general NP scenario
    ({\it{left}}) and in the NMFV scenario ({\it{right}}).
    Results from all the neutral meson systems are shown.}
  \label{fig:scales}
\end{figure}

In the case of the general NP scenario, left plot in Fig.~\ref{fig:scales} shows
the case of arbitrary NP flavour structures ($\vert F_i \vert \sim 1$) with arbitrary
phase and $L_i = 1$ corresponding to strongly-interacting and/or tree-level NP.
The overall constraint on the NP scale comes from the kaon sector (Im$\;C^4_{K}$
in Fig.~\ref{fig:scales}) and it is translated into
$\Lambda_{\mathrm{gen}} > 5.0 \cdot 10^5$~TeV.
As we are considering arbitrary NP flavour structures, the constraints on the
NP scale are very tight due to the absence of the CKM or any flavour suppression.

In the NMFV case, the strongest bound is again obtained from the kaon sector
(Im$\;C^4_{K}$ in right plot in Fig.~\ref{fig:scales})
and it translates into the weaker lower limit $\Lambda_\mathrm{NMFV} > 114$~TeV.
In this latter case and in the current scenario, the $B_s$ system also
provides quite stringent constraints.

In conclusion, a loop suppression is needed in all scenarios to obtain
NP scales that can be reached at the LHC.
For NMFV models, an $\alpha_W$ loop suppression might not be
sufficient, since the resulting NP scale is still of the order of $11$ TeV.
The general model is out of reach even for $\alpha_W$ (or stronger)
loop suppression. Finally, the reader should keep in mind the possibility
of accidental cancellations among the contribution of different operators,
that might weaken the bounds we obtained.

\bibliographystyle{JHEP}
\bibliography{bona_lhcp17_utfit}{}

\end{document}